\documentstyle[twoside,fleqn,espcrc2,epsfig]{article}

\newcommand{\AmS}{{\protect\the\textfont2
  A\kern-.1667em\lower.5ex\hbox{M}\kern-.125emS}}
\def\d{{\rm d}}
\def\Li{{\rm Li}}
\def\sab{s_{12}}
\def\sac{s_{13}}
\def\sbc{s_{23}}

\newcommand{\bubbleNLO}{
\mbox{\parbox{2.5cm}{\hspace{0.25cm}
\begin{picture}(2,1)
\thicklines
\put(0,0.5){\line(1,0){2}}
\put(1,0.5){\circle{1}}
\end{picture}
}}
\hfill}

\newcommand{\doublebubbleNLO}{
\mbox{\parbox{3.5cm}{\hspace{0.25cm}
\begin{picture}(3,1)
\thicklines
\put(0,0.5){\line(1,0){0.5}}
\put(2.5,0.5){\line(1,0){0.5}}
\put(1,0.5){\circle{1}}
\put(2,0.5){\circle{1}}
\end{picture}
}}
\hfill}

\newcommand{\triangleNLO}{
\mbox{\parbox{3cm}{\hspace{0.25cm}
\begin{picture}(2.5,1.4)
\thicklines
\put(0,0.7){\line(1,0){0.5}}
\put(1,1.2){\line(0,-1){1}}
\put(1,1.2){\line(1,0){1}}
\put(1,0.2){\line(1,0){1}}
\put(1,0.7){\circle{1}}
\end{picture}
}}
\hfill}

\newcommand{\trianglecrossNLO}{
\mbox{\parbox{3cm}{\hspace{0.25cm}
\begin{picture}(2.5,1.4)
\thicklines
\put(0,0.7){\line(1,0){0.5}}
\put(0.5,0.7){\line(1,1){0.5}}
\put(0.5,0.7){\line(1,-1){0.5}}
\put(1.5,1.2){\line(-1,-2){0.5}}
\put(1.0,1.2){\line(1,-2){0.18}}
\put(1.5,0.2){\line(-1,2){0.18}}
\put(1,1.2){\line(1,0){1}}
\put(1,0.2){\line(1,0){1}}
\end{picture}
}}
\hfill}

\newcommand{\doubleboxNLO}{
\mbox{\parbox{2.4cm}{
\begin{picture}(2.4,1.4)
\thicklines
\put(0.2,0.2){\line(1,0){2}}
\put(0.7,0.2){\line(0,1){1}}
\put(0.2,1.2){\line(1,0){2}}
\put(1.2,0.2){\line(0,1){1}}
\put(1.7,0.2){\line(0,1){1}}
\end{picture}
}}
\hfill}
\newcommand{\doubleboxdotNLO}{
\mbox{\parbox{2.4cm}{
\begin{picture}(2.4,1.4)
\thicklines
\put(0.2,0.2){\line(1,0){2}}
\put(0.7,0.2){\line(0,1){1}}
\put(0.2,1.2){\line(1,0){2}}
\put(1.2,0.2){\line(0,1){1}}
\put(1.7,0.2){\line(0,1){1}}
\put(1.2,0.7){\circle*{0.2}}
\end{picture}
}}
\hfill}

\title{
Using differential equations to compute 
two-loop box integrals\thanks{Presented at {\it Loops and Legs 
in Quantum Field Theory},
April 2000, Bastei, Germany}}

\author{\underline{T.\ Gehrmann}\address{Institut 
f\"ur Theoretische Teilchenphysik,
Universit\"at Karlsruhe, D-76128 Karlsruhe, 
Germany}
        and
        E.\ Remiddi\address{Dipartimento di Fisica,
        Universit\`{a} di Bologna, and INFN, Sezione di Bologna,
        I-40126 Bologna, Italy}}

\begin{document}
\unitlength 1cm
\begin{abstract}

The calculation of exclusive observables beyond the one-loop level
requires elaborate techniques for the computation of 
multi-leg two-loop integrals. We 
discuss how the large number of different integrals 
appearing in actual two-loop calculations can be reduced to a small
number of master integrals. An efficient method to 
compute these master integrals is to derive and solve
differential equations in the external
invariants for them. 
As an application of the differential equation method, we compute 
the ${\cal O}(\epsilon)$-term of a particular combination of 
on-shell massless planar double 
box integrals, which appears in the tensor reduction of $2 \to 2$
scattering amplitudes at two loops. 

\end{abstract}


\maketitle

\thispagestyle{myheadings}
\markright{TTP00-10 -- hep-ph/0005232}

\section{Introduction}
Perturbative corrections to many inclusive quantities have been computed 
to the two- and three-loop level in past years. From the technical 
point of view, these inclusive calculations correspond to the computation of 
multi-loop two-point functions, for which many elaborate
calculational tools have been developed. In contrast, 
corrections to exclusive
observables, such as jet production rates, could up to now only be
computed at the one-loop level. These calculations demand the 
computation of multi-leg amplitudes to the required number of loops,
which beyond the one-loop level 
turn out to be a calculational challenge obstructing further progress. 
Despite considerable progress made in recent times, many of the
two-loop integrals relevant for the calculation of jet observables
beyond next-to-leading order are still unknown. One particular class of
yet unknown
integrals appearing in the two-loop corrections to three jet production
in electron-positron collisions, to two-plus-one jet production in 
electron-proton collisions and to vector boson plus jet production in 
proton-proton collisions are two-loop four-point functions with massless 
internal propagators and one external leg off-shell. 

We elaborate on in this talk several
techniques to compute multi-leg amplitudes beyond one loop. We
demonstrate how integration-by-parts identities (already known to be a very
valuable tool in inclusive calculations) and identities following from
Lorentz-invariance (which are non-trivial only for integrals depending 
on at least 
two independent external momenta) can be used to reduce the large number 
of different integrals appearing in an actual calculation to a small 
number of master integrals. This reduction can be carried out
mechanically (by means of a small chain of 
computer programs), without
explicit reference to the actual structure of the integrals under
consideration and can also be used for the reduction of 
tensor integrals beyond one loop. 

The master integrals themselves, however, can not be computed from
these identities. We derive differential equations in the external
momenta for them. Solving these
differential equations, it is possible to compute the master integrals
without explicitly carrying out any loop integration, so that this
technique appears to be a valuable alternative to conventional
approaches for the computation of multi-loop integrals. 

We demonstrate the application of these tools on 
 several examples.

\section{Reduction to master integrals}
Any scalar massless two-loop integral can be brought into the form 
\begin{eqnarray}
\lefteqn{I(p_1,\ldots,p_n) = }\nonumber \\
& & \int \frac{\d^d k}{(2\pi)^d}\frac{\d^d l}{(2\pi)^d} 
\frac{1}{D_1^{m_1}\ldots D_{t}^{m_t}} S_1^{n_1} 
\ldots S_q^{n_q} \; ,
\label{eq:generic}
\end{eqnarray}
where the $D_i$ are massless scalar propagators, depending on $k$, $l$ and the 
external momenta $p_1,\ldots,p_n$ while $S_i$ are scalar products
of a loop momentum with an external momentum or of the two loop
momenta.  The topology (interconnection of
propagators and external momenta) of the integral is uniquely 
determined by specifying the set $(D_1,\ldots,D_t)$
of $t$ different propagators in the graph. The integral itself is then
specified by the powers $m_i$ of all propagators and by the selection 
$(S_1,\ldots,S_q)$  of scalar products and their powers $(n_1,\ldots,n_q)$ 
(all the $m_i$ are positive integers greater or equal to 1, while the 
$n_i$ are greater or equal to 0). 
Integrals of the same topology with the same dimension $r=\sum_i m_i$ 
of the denominator and same total number $s=\sum_i n_i$ of scalar products 
are denoted as a class of integrals $I_{t,r,s}$. The integration measure and 
scalar products appearing the above expression are in Minkowskian space, 
with the usual causal prescription for all propagators. The loop
integrations are carried out for arbitrary space-time dimension $d$, 
which acts as a regulator for divergencies appearing due to the
ultraviolet or infrared behaviour of the integrand (dimensional 
regularisation, \cite{dreg,hv}).  
  
The number $N(I_{t,r,s})$ of the integrals grows quickly as $r, s$ 
increase, but the integrals are related among themselves 
by various identities. 
One class of identities follows from the fact that the integral over the 
total derivative with respect to any loop momentum vanishes in
dimensional regularisation
\begin{equation}
\int \frac{\d^d k}{(2\pi)^d} \frac{\partial}{\partial k^{\mu}}
J(k,\ldots)  = 0,
\end{equation} 
where $J$ is any combination of propagators, scalar products
and loop momentum vectors. $J$ can be a vector or tensor of any rank. 
The resulting identities~\cite{hv,chet} are called integration-by-parts (IBP)
identities.

In addition to the IBP identities, one can also exploit the fact that
all integrals under consideration are Lorentz scalars (or, perhaps 
more precisely, ``$d$-rotational'' scalars) , which are
invariant under a Lorentz (or $d$-rotational) transformation of the 
external momenta~\cite{gr}. These Lorentz invariance (LI) identities 
are obtained from:
\begin{eqnarray}
\lefteqn{\left(p_1^{\nu}\frac{\partial}{\partial
    p_{1\mu}} - p_1^{\mu}\frac{\partial}{\partial
    p_{1\nu}} + \ldots\right. }\nonumber \\
&&
\left. \ldots + p_n^{\nu}\frac{\partial}{\partial
    p_{n\mu}} - p_n^{\mu}\frac{\partial}{\partial
    p_{n\nu}}\right) I(p_1,\ldots,p_n) = 0 \;. \nonumber \\ &&
\label{eq:li}
\end{eqnarray}
\markright{}

In the case of two-loop four-point functions, one has a total of 13
equations (10 IBP + 3 LI) for each integrand corresponding to an 
integral of class $I_{t,r,s}$, relating integrals of the same topology
with up to $s+1$ scalar products and $r+1$ denominators, plus integrals
of simpler topologies ({\it i.e.}~with a smaller number of different 
denominators). 
The 13 identities obtained starting from an integral $I_{t,r,s}$ do contain
integrals of the following types:
\begin{itemize}
\item $I_{t,r,s}$: the integral itself. 
\item $I_{t-1,r,s}$: simpler topology. 
\item $I_{t,r+1,s}, I_{t,r+1,s+1}$ : same topology, more complicated than
  $I_{t,r,s}$.  
\item $I_{t,r-1,s}, I_{t,r-1,s-1}$: same topology, simpler than 
   $I_{t,r,s}$.  
\end{itemize}
Quite in general, single identities of the above kind can be used 
to obtain the reduction of $I_{t,r+1,s+1}$ or $I_{t,r+1,s}$ integrals 
in terms of $I_{t,r,s}$ and simpler integrals - rather than to 
get information on the $I_{t,r,s}$ themselves. 

If one considers the set of all the identities obtained starting from 
the integrand of all the $N(I_{t,r,s})$ integrals of class $I_{t,r,s}$, 
one obtains 
$(N_{{\rm     IBP}}+ N_{{\rm LI}}) N(I_{t,r,s})$ identities 
which contain $N(I_{t,r+1,s+1})+N(I_{t,r+1,s})$ 
integrals of more complicated structure. It was first noticed by S.\
Laporta~\cite{laporta}
that with increasing $r$ and $s$
the number of identities grows faster than the number
of new unknown integrals. 
As a consequence, if for a given $t$-topology one considers the set of 
all the possible equations obtained by considering all the integrands up to 
certain values $r^*, s^*$ of $r, s$, for large enough $r^*, s^*$ 
the resulting system of equations, apparently overconstrained,
can be used for 
expressing the more complicated integrals, with greater values of $r, s$ 
in terms of simpler ones, with smaller values of $r, s$. 
An automatic procedure to preform
this reduction by means of computer algebra
is discussed in more detail in~\cite{eproc}.

For any given four-point two-loop topology, 
this procedure can result either in a reduction
 towards a small number (typically one or two) of integrals of the
topology under consideration and integrals of simpler 
topology (less different denominators), or even in a complete 
reduction of all integrals of the topology under consideration
towards integrals with simpler topology.
Left-over integrals of the topology under consideration are called 
irreducible master integrals or just 
master integrals. 

\section{Differential equations for master integrals}
\label{sec:de}
The IBP and LI identities allow to
express integrals of the form (\ref{eq:generic}) as a linear
combination of a few master integrals,  i.e.\ integrals which are 
not further reducible, but have 
to be computed by some different method. 

At present, the complete set of master integrals for massless on-shell 
two-loop box integrals is known
analytically~\cite{smirnov,tausk,smir2,dur} up to 
finite terms in $\epsilon=(4-d)/2$. For massless two-loop box integrals 
with one off-shell leg, several topologies are yet to be computed
analytically. A purely numerical approach for computing 
these integrals order by order in $\epsilon$ has recently been 
proposed by Binoth and Heinrich in~\cite{numbox}.

A method for the analytic
computation of master integrals avoiding the explicit
integration over the loop momenta is to derive differential equations in 
internal propagator masses or in external momenta for the master integral, 
and to solve these with appropriate boundary conditions. 
This method has first been suggested by Kotikov~\cite{kotikov} to relate 
loop integrals with internal masses to massless loop integrals. 
It has been elaborated in detail and generalised to differential 
equations in external momenta in~\cite{remiddi}; first 
applications were presented in~\cite{appl}.
In the case of four-point functions with one external off-shell leg
and no internal masses, one has three independent
invariants, resulting in three differential equations.

The derivatives in the invariants $s_{ij}=(p_i+p_j)^2$ 
can be expressed by derivatives in the external momenta:
\begin{eqnarray}
\sab \frac{\partial}{\partial \sab} & = & \frac{1}{2} \left( +
p_1^{\mu} \frac{\partial}{\partial p_1^{\mu}} +
p_2^{\mu} \frac{\partial}{\partial p_2^{\mu}} -
p_3^{\mu} \frac{\partial}{\partial p_3^{\mu}}\right)\;, \nonumber \\
\sac \frac{\partial}{\partial \sac} & = & \frac{1}{2} \left( +
p_1^{\mu} \frac{\partial}{\partial p_1^{\mu}} -
p_2^{\mu} \frac{\partial}{\partial p_2^{\mu}} +
p_3^{\mu} \frac{\partial}{\partial p_3^{\mu}}\right)\;,\nonumber \\
\sbc \frac{\partial}{\partial \sbc} & = & \frac{1}{2} \left( - 
p_1^{\mu} \frac{\partial}{\partial p_1^{\mu}} +
p_2^{\mu} \frac{\partial}{\partial p_2^{\mu}} +
p_3^{\mu} \frac{\partial}{\partial p_3^{\mu}}\right)\;. \nonumber\\ 
\label{eq:derivatives} 
\end{eqnarray}

It is evident that acting with the right hand  sides of (\ref{eq:derivatives}) 
on a master integral $I_{t,t,0}$
will, after interchange of 
derivative and integration, yield 
a combination of 
integrals of the same type as appearing in the IBP and LI identities for 
$I_{t,t,0}$, including integrals of type $I_{t,t+1,1}$ and 
$I_{t,t+1,0}$. Consequently, the scalar derivatives (on left hand side of 
(\ref{eq:derivatives}))
of  $I_{t,t,0}$ can be expressed by a linear combination of 
integrals up to  $I_{t,t+1,1}$ and 
$I_{t,t+1,0}$.
These can all be reduced (for topologies containing only one 
master integral) to $I_{t,t,0}$ 
and to integrals of simpler topology
by applying the IBP and LI identities. As a result, we obtain 
for the master integral $I_{t,t,0}$ 
an inhomogeneous linear
first order differential equation in each invariant. For topologies with 
more than one master integral, one finds a coupled system of 
first order differential equations.
The inhomogeneous term in these differential equations contains only 
topologies simpler than $I_{t,t,0}$, which are considered to be known 
if working in a bottom-up approach. 

The master integral $I_{t,t,0}$ is obtained by matching the general 
solution of its differential equation to an appropriate boundary
condition. Quite in general, finding a boundary condition is 
a  simpler problem than evaluating the whole
integral, since it depends on a smaller number of kinematical
variables. In some cases, the boundary condition can even be
determined from the differential equation itself. 

In writing down the differential equations for all master integrals 
appearing in the reduction of two-loop four-point functions with up to 
one off-shell leg, one finds that all boundary conditions can be 
related to the following four one-scale two-loop integrals:
\begin{eqnarray}
\bubbleNLO & = &  A_3 
\left(-s\right)^{d-3}\nonumber \\[3mm]
\doublebubbleNLO 
& = & A_{2,LO}^2 \left(-s\right)^{d-4}\nonumber\\[3mm]
\triangleNLO
 & = & A_4 \left(-s\right)^{d-4}\nonumber\\[3mm]
\trianglecrossNLO
 & = & A_6 \left(-s\right)^{d-6}\nonumber\\
\end{eqnarray}
These integrals fulfill homogeneous differential equations in $s$; their 
normalisation can therefore not be calculated in the differential 
equation method. They can however be evaluated explicitly 
using Feynman parameters~\cite{kl}.

\section{Applications}

Using the differential equation method, we computed in~\cite{gr} all
master integrals with up to $t=5$ different denominators that can appear 
in the reduction of two-loop four-point functions with one off-shell
leg. 

Differential equations were also used in the computation of two-loop
on-shell master integrals with planar~\cite{smir2} and
crossed~\cite{dur} topologies. These topologies contain two master
integrals each. The differential equations were applied
only to obtain a
relation among both master integrals, while one of the master integrals
was calculated by Smirnov for the planar case~\cite{smirnov} and 
Tausk for the non-planar case~\cite{tausk} using a different method.

At this workshop, Nigel Glover reported first progress towards the
computation of massless on-shell $2\to 2$ scattering amplitudes at two
loops~\cite{glover}. In the reduction of planar amplitudes, it turned
out that it is not sufficient to know the planar double box integrals 
up to their finite terms in $\epsilon$. 
In the tensor reduction procedure,
the following combination of master integrals arises:
\begin{eqnarray}
\lefteqn{\frac{1}{\epsilon}\left[ 
\doubleboxNLO - t \doubleboxdotNLO \right]}\nonumber \\
& \sim & \frac{1}{\epsilon}\left[
K_1(x,\epsilon) + K_2(x,\epsilon)\right]\;,
\end{eqnarray}
where $x=t/s$. The functions $K_1(x,\epsilon)$ and $ K_2(x,\epsilon)$
were computed only up to finite terms in $\epsilon$ by Smirnov and 
Veretin in~\cite{smir2}. The structure of the up to now unknown 
${\cal O}(\epsilon)$-term was subject of debate at this workshop. We
did therefore decide after the workshop
to attempt its computation using the 
differential equation method. 

Starting from the differential equations for the massless double box 
integrals with one off-shell leg, which were obtained using the
algorithms described in Section~\ref{sec:de}, we can derive the 
differential equations for the above on-shell integrals 
by constraining $\sab+\sac+\sbc =0$ (see also~\cite{dur}). Using $s$ and 
$x$ as independent variables, we obtain two homogeneous differential 
equations in $s$ (corresponding to a rescaling relation,~\cite{appl}) and 
two coupled 
inhomogeneous equations in $x$, which can be employed to compute the 
two master integrals. Using the fact that the master integrals and their 
derivatives are regular in $t$ at 
$t=-s$, the boundary condition at $x=-1$
is inferred from the differential equations. This condition is
however not yet sufficient to 
determine the boundary conditions for both master 
integrals, since it only fixes one combination of them. 
It does however turn out that it is sufficient to match the
non-logarithmic terms in $K_1(x,\epsilon)$, obtained by Smirnov 
in~\cite{smirnov},  up to finite order in 
$\epsilon$ to find the boundary condition 
for $K_1(x,\epsilon)+K_2(x,\epsilon)$ up to ${\cal O}(\epsilon)$. 
The differential equations are then solved order by order in $\epsilon$
by expressing the unknown as
 sum of harmonic polylogarithms~\cite{hpl}. 
Requiring the differential equation to be fulfilled and the 
boundary conditions to be matched, the coefficients of the
harmonic polylogarithms in the ansatz can all be determined. 

Using the same normalisation factor as~\cite{smir2}, 
the coefficient of the ${\cal  O}(\epsilon)$-term of the above
combination of double box integrals reads:
\begin{eqnarray}
\lefteqn{\left[
K_1(x,\epsilon) + K_2(x,\epsilon)\right] \Big|_{{\cal 
O}(\epsilon)} }\nonumber \\
& = & -\frac{4}{3} \ln^4 x + \frac{4}{3} \ln^3 x - 4(4+3\pi^2) \ln^2 x
\nonumber \\
& & -\Big(46 + 21 \pi^2 - \frac{16}{3} \zeta (3)\Big) \ln x \nonumber \\
& &  - 50 - 8\pi^2 - \frac{139}{60}\pi^4 + \frac{698}{3}\zeta(3)
 \nonumber\\
& & -32\, S_{2,2}(-x) + 32 \ln x\, S_{1,2}(-x) \nonumber \\
& & - 128\, \Li_4(-x)
+32 \Big( 3\ln x - \ln (1+x) \nonumber \\
& & -2\Big)\, \Li_3(-x)+32\Big( -\ln^2 x + 2 \ln x   \nonumber \\
& & + \ln x \ln (1+x) - \frac{5}{6} \pi^2 \Big)
 \Li_2(-x) \nonumber \\
& & + 8 \left(\ln^2 x + \pi^2 \right)\ln^2(1+x) \nonumber\\
& & + \frac{16}{3} \Big( -\ln^3 x + 6 \ln^2 x + 6 \pi^2 - 2 \pi^2 \ln x 
\nonumber \\
&& + 6 \zeta(3) \Big) \ln(1+x) \nonumber \\
&& + x \bigg[
 +\frac{16}{3} \ln^4 x - \frac{52}{3} \ln^3 x +\frac{46}{3}\pi^2 \ln^2 x
\nonumber \\
&& + \Big(2 -\frac{169}{3} \pi^2 - \frac{496}{3} \zeta (3)\Big) \ln x 
\nonumber \\
& &  - 84 - \frac{46}{3}\pi^2 + \frac{823}{360}\pi^4 + 536\,\zeta(3)
 \nonumber\\
& & -56\, S_{2,2}(-x) + 56 \ln x\, S_{1,2}(-x)  \nonumber \\
&& + 40 \, \Li_4(-x)
+ \Big( 16 \ln x - 56 \ln (1+x) \nonumber \\
&& - 104\Big)\, \Li_3(-x) - \Big( 36 \ln^2 x +\frac{8}{3}\pi^2 \nonumber \\
&&
- 56 \ln x \ln(1+x) - 104 \ln x 
\Big)  \Li_2(-x) \nonumber \\
&& + 14 (\ln^2 x + \pi^2) \ln^2 (1+x) \nonumber \\
&& + \frac{4}{3} \Big(-16 \ln^3 x + 39 \ln^2 x - 23\pi^2 \ln x \nonumber \\
&&
+ 39 \pi^2 +
42 \zeta(3) \Big) \ln(1+x) \bigg] \;.
\end{eqnarray}

\section{Outlook}

We have demonstrated how techniques developed for 
multi-loop calculation of two-point functions can be extended towards 
integrals with a larger number of external legs. In particular, we 
have shown that the use of differential equations in external invariants 
can be used as a powerful method to compute master integrals without 
carrying out explicit loop integrations.
As a first example of 
the application of these tools in practice, we computed some 
up to now unknown 
two-loop four-point functions, relevant for jet calculus beyond the 
next-to-leading order. 
 The most important potential application of these 
tools is the yet outstanding derivation of two-loop virtual corrections to 
exclusive quantities, such as jet observables.

\end{document}